\documentclass[prl,9pt,twocolumn]{revtex4-1}

\usepackage[utf8]{inputenc}

\usepackage{graphicx}

\usepackage{verbatim}

\usepackage{amsmath}
\usepackage{amssymb}
\usepackage{bm}
\usepackage{braket}

\begin{document}
	\title{Self-calibrating Optical Low-Coherence Reflectometry with Energy-Time Entangled Photons for Absolute Distance Measurements}
	
	\author{Manuel Unternährer, André Stefanov}

\affiliation{Institute of Applied Physics, University of Bern, 3012 Bern, Switzerland}

\begin{abstract}
	Optical low-coherence reflectometry is capable of unambiguously measuring positions of stacked, partially reflective layers in a sample object. It relies on the low coherence of the light source and the absolute distances are obtained from the position reading of a mechanical motor stage. We show how to exploit the simultaneous high and low coherence properties of energy-time entangled photon pairs to directly calibrates the position scale of an OLCR scan with a reference laser wavelength. In experiment, a precision of 1.6\,nm and good linearity is demonstrated.
\end{abstract}
\maketitle
\section{Introduction}

Through the definition of its speed, light is the primary tool for absolute length measurements. Direct measurements by time of flight of pulsed lasers are in practice limited by the temporal resolution of detectors to the millimeter range \cite{Amann2001}. On the other side, the most precise length determinations are obtained by interferometric phase measurements with continuous light \cite{Yoshizawa2015}. However distances are in that case only determined with an ambiguity corresponding to the wavelength of light, usually in the sub-micrometer range. Beside the practical limits of detectors, it is the spectrum of light that ultimately determines the achievable uncertainty on the length and the unambiguous range of measurement. Between two extreme cases, monochromatic light and ultra-short laser pulses,  intermediate regimes have  been developed that allows to achieve good resolution with reduced ambiguity. For instance multi-wavelength interferometry or interferometric measurements with dual femtosecond optical frequency combs \cite{Coddington2009,Lee2013a}. Another intermediate regime is accessible by optical low-coherence reflectometry (OLCR), also called optical coherence tomography (OCT), where distances can be measured over long range with micrometer accuracy. It exploits coherence properties of light to measure the positions of a sample's partially reflective layers with resolution in axial direction. It has become standard tool in medicine and biomedical applications \cite{Zysk2007}. In classical optical length metrology, the coherence of light and its spectral density are the relevant quantities, that determines the achievable precision and ranges. For instance in OLCR, the resolution is essentially given by the coherence length of the source, that is inversely proportional to its spectral bandwidth; while in dual combs schemes, the key element is to exploit light sources that have both narrow and broad spectral features.

While coherent classical light is fully described by its one dimensional complex spectrum, quantum light sources offer a much richer space of parameters to play with. For instance Abbourady et al. showed in theory and experiment that the transition from a classical broadband light source to a quantum light can be beneficial for OLCR \cite{Abouraddy2002,Nasr2003}. In particular, the frequency correlation from entangled photon pairs allows for the compensation of the sample's chromatic dispersion and thereby undisturbed axial resolution in quantum OCT (QOCT). While in that case the emission geometry was non-collinear, i.e. the photons are emitted in separate spatial directions, it was later shown that the same signal can be recovered in a collinear system \cite{Lopez-Mago2012}.

The spectrum of frequency-entangled photon pairs is described by a two-dimensional function, that characterizes the correlations between the frequencies of the photons. Hence, the behavior of such light differs if one observe one photon only, or two photons in coincidence. In this work, we exploit for precise length measurements the simultaneous presence of two types of interferences in the QOCT signal: Low coherence in one-photon counting and high coherence in two-photon coincidence. The first ones provide an unambiguous long range measurement, similar to the interferometer  signal with a classical low coherence source, while the second one allows for interferometric precision relative to a fixed laser wavelength. It is visible over the full scanning range and its well-defined oscillation period allows to establish a position scale independent of the delay stage position feedback. This self-calibrating feature permits to measure distances between surfaces in terms of multiples of the wavelength of a narrowband laser. Sub-wavelength precision can be achieved by appropriate signal processing. In our experimental implementation, the measurement of a distance of 0.28~mm between two mirrors is shown to be reproducible within 1.6~nm standard deviation. Furthermore, this measurement method is verified to have a nearly perfect linearity over different magnitudes of distance ranges.

\begin{figure}
	\centering
	\includegraphics[scale=1]{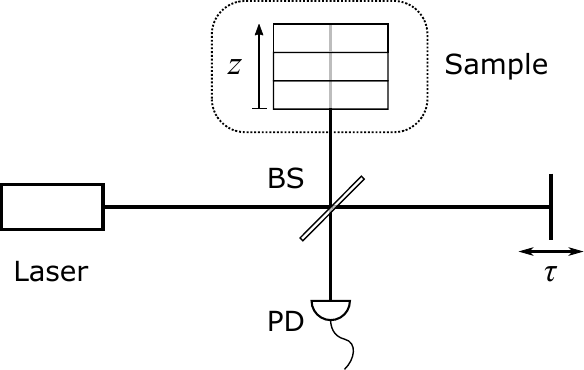}
	\caption{Classical OLCR setup. The  Michelson interferometer consists of a beamsplitter (BS), a reference arm with delay $\tau$ and a sample arm with multiple reflective layers. A broadband laser serves as illumination. The interference signal is recorded by a photodiode (PD).}
	\label{fig:OLCRsetup}
\end{figure}

\section{Theory}

\subsection{Classical OLCR}

A classical OLCR setup is depicted in Fig.~\ref{fig:OLCRsetup}. The sample object is placed in one arm of a Michelson interferometer illuminated by a classical broadband light source. The interferometer output shows interference oscillations if the arms are balanced within the short coherence time of the source. Therefore, the relative position between different surfaces can be determined within an error given by the accuracy of the delay stage position feedback in the scanning arm. More specifically, displacing the reference arm mirror position by $d$ introduces a temporal delay $\tau=2d/c$ with $c$ being the speed of light. The intensity as a function of the delay $I(\tau)$ at the interferometer output is recorded with a photodiode (PD).

A transfer function $ H(\omega)$ characterizes the reflectivity properties of a sample object. In the case of $n$ stacked and partially reflecting surfaces, neglecting multiple reflections, it reads
\begin{equation}\label{eq:H}
 H(\omega) = \sum_{j=1}^n r_j \, e^{i  \omega \tau_j}
\end{equation}
with  reflection amplitudes $r_j\in[0,1]$ and introduced time delays $\tau_j$  \cite{Abouraddy2002}. Assuming free-space propagation without optically dense or dispersive media between surfaces, the temporal delays are related to the sample surface positions $z_j$ by $\tau_j = {2z_j}/{c}$.

A broadband light source with central frequency $\omega_0$ is characterized by the spectral power density  $S(\Omega)$ which is defined in terms of the relative frequency $\Omega := \omega - \omega_0$. The intensity measured at the interferometer output is
$ I(\tau) = \Gamma_0 + 2 \,\text{Re}\left\{\Gamma(\tau) \, e^{-i\omega_0\tau}\right\} $
depending on the time delay $\tau$ introduced in the reference arm \cite{Abouraddy2002}.
$\Gamma_0$ is a constant while the cross-interference between sample and reference arm shows oscillations with the envelope
$ \Gamma(\tau) = \int d\Omega\, H(\omega_0+\Omega) S(\Omega) e^{-i\Omega\tau}$.
Using \eqref{eq:H} and surface separations larger than the coherence time of the light source, we can rewrite
\begin{equation}\label{eq:Itau}
I(\tau)  = I_0 +  \sum_{j=1}^n  r_j\, f(\tau-\tau_j)
\end{equation}
with the single surface interferometer response function $ f(\tau) = 2\, \text{Re} \{ s(\tau) \,e^{-i\omega_0\tau}\}$
where the envelope $s(\tau)$ is the inverse Fourier transform of $S(\Omega)$ . The even function $f(\tau)$ is therefore the light source's electric field temporal auto-correlation. It is maximal at $\tau = 0$ and has a width given by the coherence time of the source. A broad spectrum (low coherence time) yields narrow interferometer oscillations centered around every surface position.

In order to reconstruct the surface positions $z_j$ from the measurement of $I(\tau)$, a digital signal auto-correlation can be performed in a post-processing step with
\begin{align}
A(\Delta\tau) &= \int d\tau\, I(\tau)I(\tau+\Delta\tau)  \nonumber \\
&=  A_0 + \sum_{i,j} r_i r_j \int d\tau\, f(\tau)f(\tau+\Delta\tau+\tau_j-\tau_i). \label{eq:Autocorr}
\end{align}
It shows very distinct peaks at every $ \Delta\tau = \tau_i - \tau_j$ due to the fact that $f(\tau)$ has a narrow envelope given by the coherence time and that it is oscillatory at a period of one wavelength. The evaluation of the peaks of this auto-correlation determine the position differences at sub-wavelength precision.

\subsection{Quantum OLCR }

 \begin{figure}
	\centering
	\includegraphics[scale=0.8]{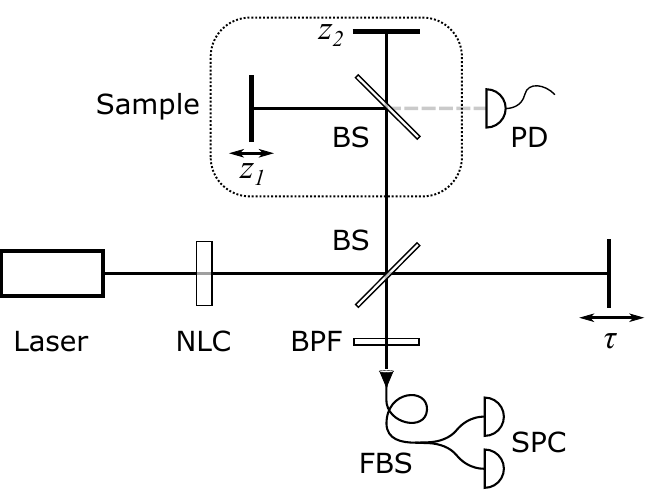}
	\caption{Quantum OLCR setup. SPDC light generated in a CW pumped non-linear crystal (NLC) is inserted into a Michelson interferometer consisting of: Beamsplitter (BS),  bandpass filter transmitting SPDC emission (BPF), fiber beamsplitter (FBS), single photon counters (SPC) and  scanning reference arm with delay $\tau$. The test sample consists of two surfaces at positions $z_1$ and $z_2$, whose relative distance is interferometrically locked to the pump wavelength. }
	\label{fig:QOLCRsetup}
\end{figure}

\begin{figure*}
	\centering
	\includegraphics[scale=0.7, trim={2cm 0 2cm .0cm}]{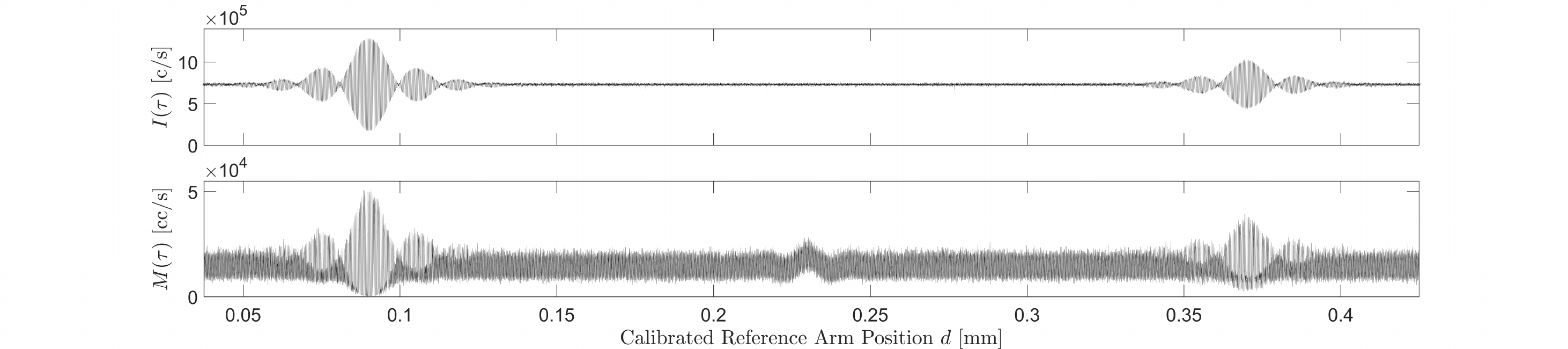}
	\caption{QOLCR scan with intensity  (top) and simultaneous coincidence  (bottom) measurements. In intensity $I(\tau)$, the two sample surfaces produce interference fringes only for reference arm positions $d$ in their vicinity. The coincidence signal $M(\tau)$  shows two-photon interference over the full scanning range. The  position scale $d = c\tau/2$ is calibrated using this sinusoidal signal.}
	\label{fig:G1G2}
\end{figure*}

In quantum optical low-coherence reflectometry (QOLCR), a photon pair source using type-0 spontaneous parametric down-process (SPDC) \cite{Hong1985} is used instead of a broadband laser. A narrowband laser emitting at frequency $\omega_p$ pumps a non-linear crystal. 
In the approximation of a perfectly monochromatic pump laser, the joint quantum state of the generated photon pair is given by
$$
 \ket{\Psi} = \int d\Omega \, \Lambda(\Omega) \ket{\omega_0+\Omega, \omega_0-\Omega} 
 $$
where $\Omega := \omega - \omega_0$ is the relative frequency with respect to the central emission frequency $\omega_0 = \omega_p / 2$. 
The two-photon spectral amplitude $\Lambda(\Omega)$ defines the spectral density $S(\Omega) = |\Lambda(\Omega)|^2$. Both photons are assumed to be emitted into the same polarization and spatial (detection) mode. 
With SPDC as the light source in the Michelson interferometer, an intensity measurement $I(\tau)$ at the output port yields the classical result  of \eqref{eq:Itau} with the SPDC spectrum $S(\Omega)$ used. 
  
In a second-order correlation measurement, new features arise. As derived in \cite{Lopez-Mago2012}, the two-photon coincidence rate measured at the output port, using a beam splitter and two detectors, is given by
\begin{multline}\label{eq:M}
 M(\tau) = M_0 + 2\, \text{Re}\{M_1(2\tau)\}  \\ 
 + 4 \,\text{Re}\{M_2(\tau)\,  e^{-i\omega_0\tau}\} + 2 \, \text{Re}\{M_3 \, e^{-i2\omega_0\tau}\}
\end{multline}
at a reference arm delay $\tau$. The interference terms $M_1(2\tau)$ and $M_2(\tau)$, corresponding to Hong-Ou-Mandel and single-photon interference, are slowly varying at the time scale of the first-order coherence time corresponding to the width of the Fourier transform of the spectrum $S(\Omega)$. The last term describes two-photon interference (TPI) and is sinusoidally oscillating at double central frequency $\omega_0$ with a constant amplitude $M_3$. This term is originating from the interference of amplitudes where both photons take either the reference or the sample arm and is given by $M_3 = \int d\Omega\, H(\omega_0+\Omega)\, H(\omega_0-\Omega) \, S(\Omega)$ \cite{Lopez-Mago2012}. Inserting $H(\omega)$ from \eqref{eq:H} for the case of reflecting surfaces which are separated much more than the two-photon correlation time, this yields $ M_3 = S_0 \, \sum_{j=1}^n r_j^2 \, e^{-i2\omega_0\tau_j}$  with total light source power $S_0$. Therefore, $M_3$  is in general a non-zero constant which only for very specific reflection coefficients $r_j$ and separations $\tau_j$ vanishes.  
In contrast to the first-order interference fringes in $I(\tau)$ which are only visible  close to a sample surface position, 
the TPI oscillations in $M(\tau)$ are present at constant amplitude over the full measurement range of $\tau$ due to the long two-photon coherence time inherited from the pump laser.

\subsection{Self-Calibration Signal Processing in QOLCR}

In a interferometer scan, the real value $\tau$ at a measurement point is not known, but an approximate value  $\tau'$ is assumed to be provided by the measurement apparatus. For instance, it can be measured by a coarse motor encoder of the reference arm mirror position $d$ by $\tau'=2d/c$, or in a fixed motor speed setting by $\tau' = 2vt/c$ with time $t$ and velocity $v$.

In a QOLCR measurement,  coincidences $M(\tau')$ are being measured simultaneously with $I(\tau')$ while the reference arm is scanned over the measurement range of $\tau'$. As obvious from  \eqref{eq:M}, the TPI is well separated from the others in terms of oscillation frequency. It can be extracted from $M(\tau')$ in a post-processing step by applying a digital high-pass filter \cite{Lopez-Mago2012}. Its very well defined and narrow oscillation frequency $2\omega_0 = \omega_p$ serves as a position reference signal. 

By knowing the exact value of $\omega_p=2\pi c/\lambda_p$, the extracted signal $\text{TPI}(\tau') = 2\,\text{Re}\{M_3 \,e^{i\,\omega_p\tau(\tau')}\}$ allows with a fitting or phase extraction procedure (e.g. using peak finding and interpolation) to get the relation $ \tau' = \tau'(\tau)$ between the exact $\tau$ value and its coarse measurement $\tau'$. From the measured classical intensity signal $I(\tau')$, a calibrated signal can be calculated with
$$ 
	I_c(\tau) := I(\tau'(\tau)).
$$
By interpolation, a linearly spaced $\tau$ scale can be established. Applying the  auto-correlation procedure of \eqref{eq:Autocorr} on $I_c(\tau)$, measurements of the surface distances are independent of the approximate length measurement but are directly linked to the known wavelength of the pump laser.

\section{Experiment}

The QOLCR setup is shown if Fig.~\ref{fig:QOLCRsetup}. A non-linear, periodically poled KTP crystal serves as type-0 SPDC light source. It is pump by a grating stabilized diode laser of 30\,mW power at $\lambda_p=405$\,nm with less than 2\,MHz bandwidth and a collimated beam of 1\,mm radius. The laser wavelength is locked to an etalon cavity and guarantees a  long-time wavelength stability. SPDC phase matching is chosen for nearly frequency-degenerated, collinear emission where photon pairs centered around $\lambda_0=810$\,nm are emitted and can be coupled into the same spatial detection mode.

In the reference arm, a standard motorized translation stage (Thorlabs PT1-Z8) is used.
The stage is driven at constant speed of 500\,nm/s while the detectors measure continuously. The detection uses a single mode fiber with an adjustable collimator. The detection mode is therefore nearly Gaussian and its waist of 0.2\,mm radius is positioned at the crystal center and aligned to the pump beam for maximal coincidence signal \cite{Guerreiro2014}.  The power spectrum of the coupled SPDC photon pairs can be measured interferometrically and show 30\,nm bandwidth. The used reference arm scanning range of 0.3\,mm introduces in the detection mode a negligible Gouy phase corresponding to 0.12\,nm position shift. 
A 50:50 fiber beam splitter distributes the interferometer output to two fiber-coupled single photon counting modules. An electric coincidence circuit with 10\,ns coincidence window is used to detect photon pairs. The detection event counters are read at a rate of 100\,Hz corresponding to a resolution of 5\,nm in delay arm position at the mentioned velocity.

For testing the QOLCR method, a sample consisting of two reflecting surfaces is used. As depicted in Fig.~\ref{fig:QOLCRsetup}, its surfaces are mirrors behind the two output ports of a beamsplitter. One mirror is fixed at distance $z_2$, the other at $z_1$ is on a nano-positioner with sub-nanometer accuracy (MCL Nano-OP30). Using the pump light incident into the sample arm, the relative position between the mirrors $z_1-z_2$ is interferometrically locked to the pump wavelength using the feedback from the photodiode (PD) measuring  pump light in the sample arm.

\section{Results}

\begin{figure}
\centering
\includegraphics[scale=0.25,trim={0.0cm 0 0cm 0cm},clip]{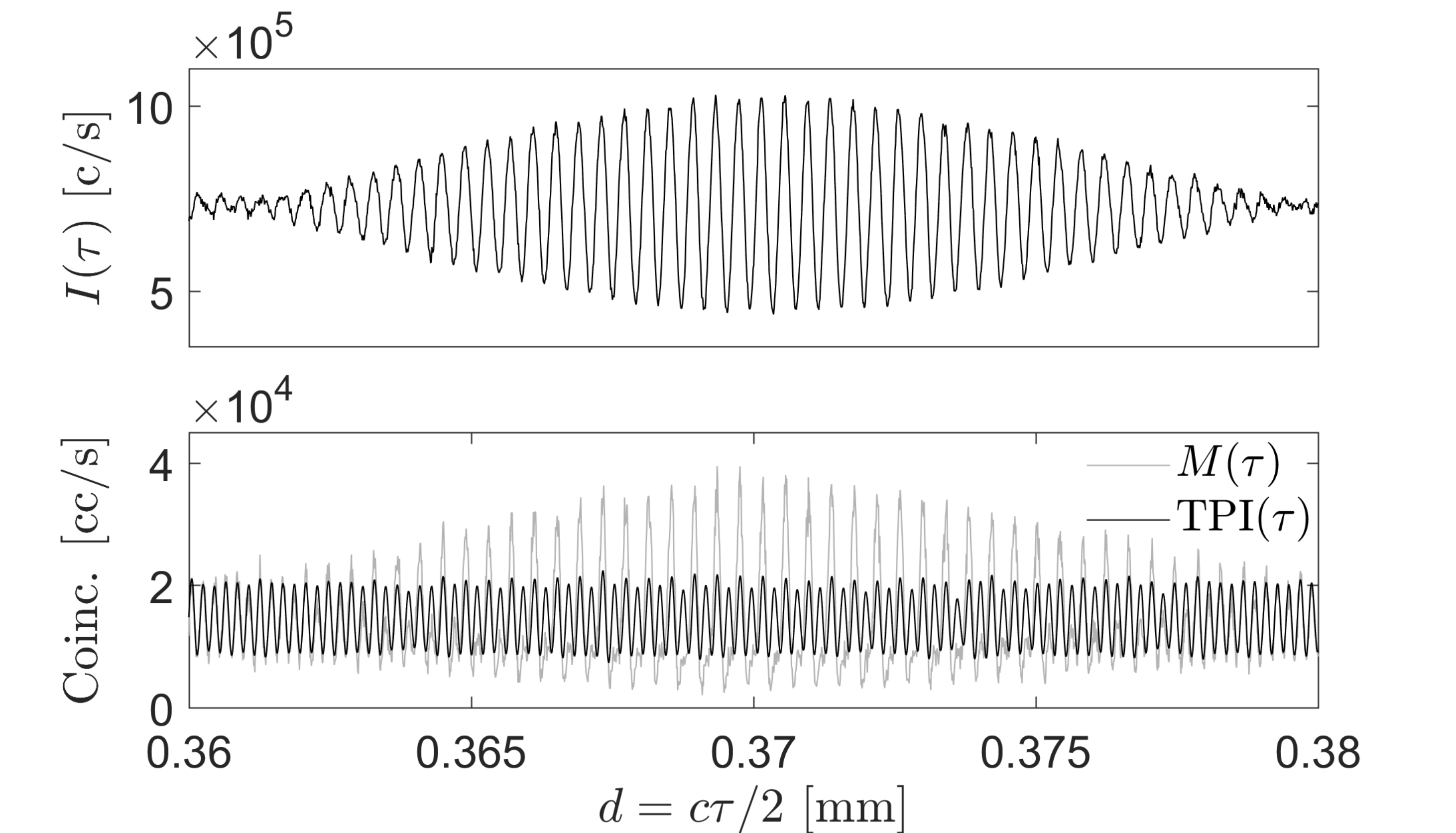}

\caption{QOLCR scan with intensity (top) and coincidence (bottom) measurements corresponding to a magnified region of Fig.~\ref{fig:G1G2}. Even in the presence of intensity fringes and single-photon interference in the signal $M(\tau)$, a sinusoidal $\text{TPI}(\tau)$ can be extracted by band-pass filtering the digital signal $M(\tau)$.}
\label{fig:G1G2zoom}
\end{figure}

\begin{figure}
\centering
\includegraphics[scale=0.45,trim={0cm 0 0cm 0cm},clip]{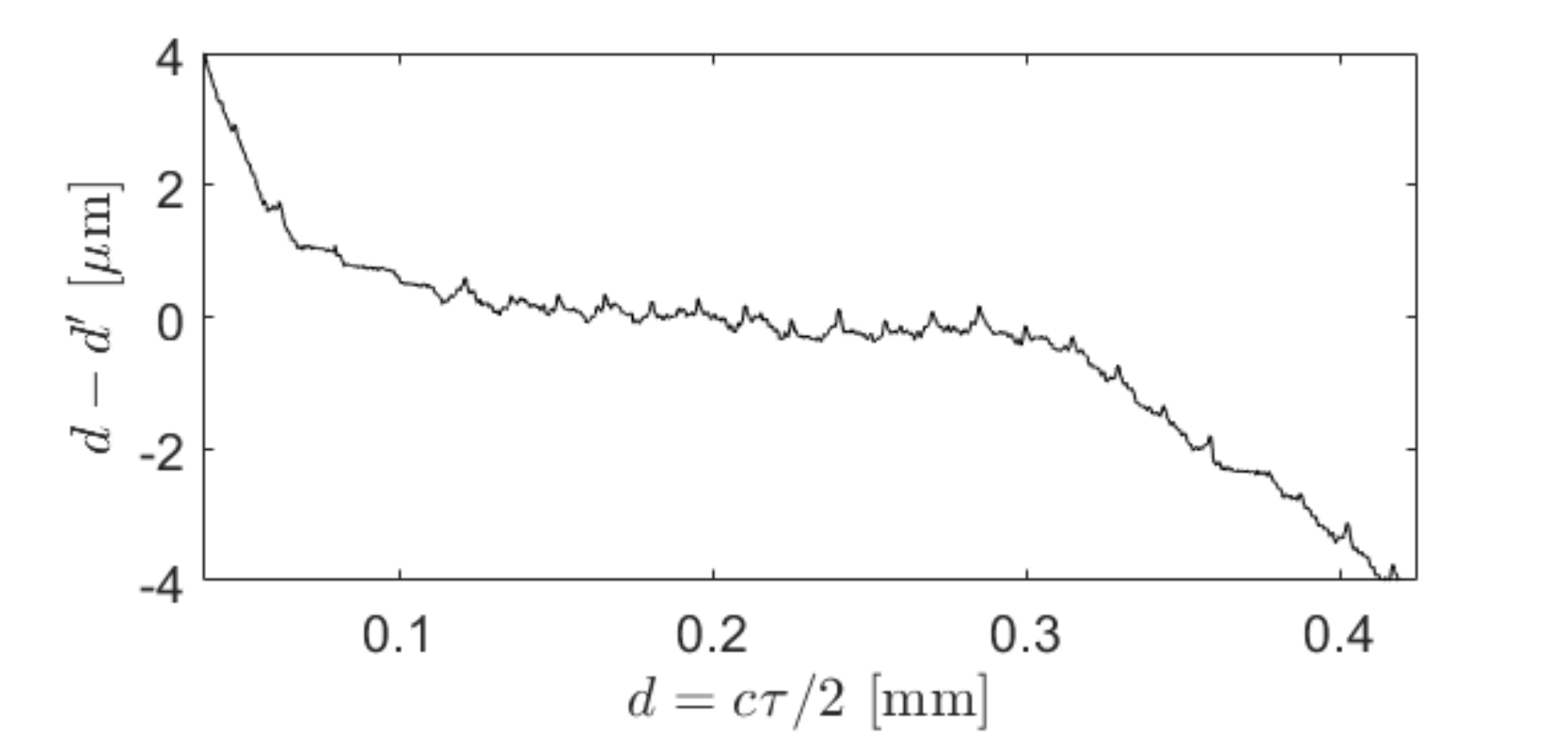}
\caption{ Correction of the self-calibrated position $d$ to the approximate position  $d'$ from the motor encoder. These deviations are due to mechanical imprecision of the motor and translation stage.}
\label{fig:dCorrection}
\end{figure}

A continous OLCR scan of the reference arm length $d$ is performed while measuring intensity $I(\tau)$ and coincidence $M(\tau)$, see Fig.~\ref{fig:G1G2}. The two fringe envelopes of $I(\tau)$ correspond to the surfaces of the sample while no interference is visible in between.  $M(\tau)$ shows single-photon interference fringes at the surface positions, a Hong-Ou-Mandel interference feature centered between them, and the TPI signal which is present in all regions. By applying a digital band-pass filter, single-photon interference can be removed and TPI recovered as shown in the magnified region of Fig.~\ref{fig:G1G2zoom}. Minor distortions of the TPI are present as the single-photon interference signal is not fully suppressed by the filter. The TPI is used for determining the calibrated reference arm position $d$. The correction of this calibrated scale to the approximate motor stage encoder is shown in Fig.~\ref{fig:dCorrection}.

The intensity signal auto-correlation $A(\tau)$ from \eqref{eq:Autocorr} can now be analyzed with the calibrated scale. A parabolic fit of its envelope allows to  identify the central peak corresponding to the distance $z_1-z_2= 0.280228$\,mm. In order to quantify the measurement precision, 70 successive measurements are performed for this fixed sample size. A measurement value variation corresponding to a standard deviation of 1.6\,nm is determined. Four outliers at one wavelength shift, due to misidentification of the auto-correlation peak, are ignored in this analysis.

\begin{figure}
\centering
\includegraphics[scale=0.35,trim={0.3cm 3.6cm -1cm 2.5cm},clip]{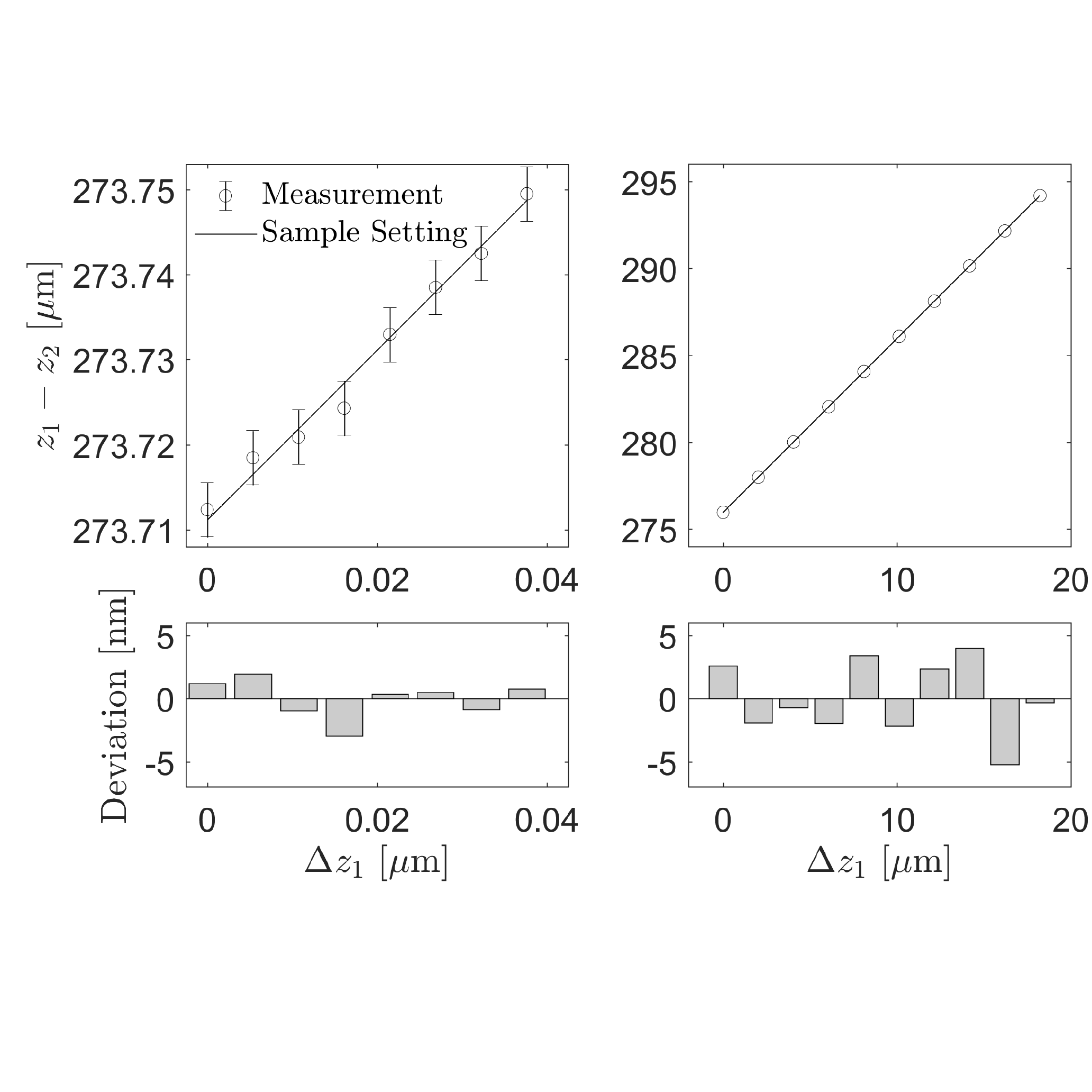}
\caption{Linearity analysis of the QOLCR method. Top row: The distance $z_1-z_2$ is measured for different sample settings where $z_1$ is varied in steps of 5\,nm (left) and 2025\,nm (right). The expected linearity of unit slope is shown (lines) with measurement and $2\sigma$ errors. Bottom row: Corresponding deviations of the measurements from the expected values on a nanometer scale.}
\label{fig:dLinearity}
\end{figure}

Changing the sample surface distance by varying of the position $z_1$ using the nano-positioner allows to study the linearity of the measurement method. Fig.~\ref{fig:dLinearity} shows the step-wise increase of $z_1$ where single QOLCR measurements are performed.  With step sizes of 5\,nm and 2025\,nm, we observe very good linearity over the full range with deviations of less than 6\,nm from ideal values indicated by lines of unit slope. The 5\,nm steps are realized by changing the set-point of the sample control loop according to the theoretical response of its interferometric feedback signal. The larger step is a multiple of the oscillation period $\lambda_p/2$ and use the same, constant controller set-point after the nano-positioner performed the desired step.

\section{Conclusion}

Frequency-entangled SPDC light exhibits simultaneous narrow- and broadband features. We exploit this property to combine both the advantages of broadband classical OLCR for unambiguous distance measurements and of monochromatic light for relating distances to the wavelength of the laser, that can be in principle directly traced back to primary length standards.
The QOLCR experiment shows very high precision for length measurements. The tolerance of the method against imperfections of the delay stage was successfully demonstrated.  Despite the wavelength of 810\,nm and only one measurement per 5\,nm, a measurement precision of 1.6\,nm was achieved. This can be understood in terms of the data interpolation and implicit averaging happening in the auto-correlation approach for determining the distances. This method also shows high robustness against uncorrelated noise in the intensity measurement due to its averaging.

Finally, as in many proposed optical quantum metrology schemes, the question of the relevance of entanglement arises \cite{Stefanov2017}. It can be shown that QOLCR and the proposed quantum OCT methods \cite{Teich2012} can be mimicked by an appropriate classical light source. It requires the emission of two narrow frequencies which are randomly varying but anti-correlated around $\omega_0$, while the detection is performed by two photodiodes which measure the output intensity for $\omega>\omega_0$ and $\omega<\omega_0$ separately. The product of the intensities would show an interference pattern identical to TPI \cite{Mazurek2013,Shirai2014a}. Therefore, it is not entanglement per se that provides advantages in such schemes, but the presence of energy correlations. Nevertheless, SPDC light offers the additional advantage to be very robust against background light: The two-photon interference signal is temporally strongly correlated and is thereby well distinguishable from uncorrelated background.

\section*{Funding Information}
This research was supported by the grant PP00P2\_159259 of the Swiss National Science Foundation.


\begin{thebibliography}{14}%
	\makeatletter
	\providecommand \@ifxundefined [1]{%
		\@ifx{#1\undefined}
	}%
	\providecommand \@ifnum [1]{%
		\ifnum #1\expandafter \@firstoftwo
		\else \expandafter \@secondoftwo
		\fi
	}%
	\providecommand \@ifx [1]{%
		\ifx #1\expandafter \@firstoftwo
		\else \expandafter \@secondoftwo
		\fi
	}%
	\providecommand \natexlab [1]{#1}%
	\providecommand \enquote  [1]{``#1''}%
	\providecommand \bibnamefont  [1]{#1}%
	\providecommand \bibfnamefont [1]{#1}%
	\providecommand \citenamefont [1]{#1}%
	\providecommand \href@noop [0]{\@secondoftwo}%
	\providecommand \href [0]{\begingroup \@sanitize@url \@href}%
	\providecommand \@href[1]{\@@startlink{#1}\@@href}%
	\providecommand \@@href[1]{\endgroup#1\@@endlink}%
	\providecommand \@sanitize@url [0]{\catcode `\\12\catcode `\$12\catcode
		`\&12\catcode `\#12\catcode `\^12\catcode `\_12\catcode `\%12\relax}%
	\providecommand \@@startlink[1]{}%
	\providecommand \@@endlink[0]{}%
	\providecommand \url  [0]{\begingroup\@sanitize@url \@url }%
	\providecommand \@url [1]{\endgroup\@href {#1}{\urlprefix }}%
	\providecommand \urlprefix  [0]{URL }%
	\providecommand \Eprint [0]{\href }%
	\providecommand \doibase [0]{http://dx.doi.org/}%
	\providecommand \selectlanguage [0]{\@gobble}%
	\providecommand \bibinfo  [0]{\@secondoftwo}%
	\providecommand \bibfield  [0]{\@secondoftwo}%
	\providecommand \translation [1]{[#1]}%
	\providecommand \BibitemOpen [0]{}%
	\providecommand \bibitemStop [0]{}%
	\providecommand \bibitemNoStop [0]{.\EOS\space}%
	\providecommand \EOS [0]{\spacefactor3000\relax}%
	\providecommand \BibitemShut  [1]{\csname bibitem#1\endcsname}%
	\let\auto@bib@innerbib\@empty
	\bibitem [{\citenamefont {Amann}\ \emph {et~al.}(2001)\citenamefont {Amann},
		\citenamefont {Bosch}, \citenamefont {Lescure}, \citenamefont {Myllyla},\
		and\ \citenamefont {Rioux}}]{Amann2001}%
	\BibitemOpen
	\bibfield  {author} {\bibinfo {author} {\bibfnamefont {M.-C.}\ \bibnamefont
			{Amann}}, \bibinfo {author} {\bibfnamefont {T.}~\bibnamefont {Bosch}},
		\bibinfo {author} {\bibfnamefont {M.}~\bibnamefont {Lescure}}, \bibinfo
		{author} {\bibfnamefont {R.}~\bibnamefont {Myllyla}}, \ and\ \bibinfo
		{author} {\bibfnamefont {M.}~\bibnamefont {Rioux}},\ }\href {\doibase
		10.1117/1.1330700} {\bibfield  {journal} {\bibinfo  {journal} {Optical
				Engineering}\ }\textbf {\bibinfo {volume} {40}},\ \bibinfo {pages} {10}
		(\bibinfo {year} {2001})}\BibitemShut {NoStop}%
	\bibitem [{\citenamefont {Yoshizawa}(2015)}]{Yoshizawa2015}%
	\BibitemOpen
	\bibinfo {editor} {\bibfnamefont {T.}~\bibnamefont {Yoshizawa}},\ ed.,\
	\href@noop {} {\emph {\bibinfo {title} {{Handbook of OPTICAL METROLOGY,
					Principles and Applications}}}}\ (\bibinfo  {publisher} {CRC Press},\
	\bibinfo {year} {2015})\BibitemShut {NoStop}%
	\bibitem [{\citenamefont {Coddington}\ \emph {et~al.}(2009)\citenamefont
		{Coddington}, \citenamefont {Swann}, \citenamefont {Nenadovic},\ and\
		\citenamefont {Newbury}}]{Coddington2009}%
	\BibitemOpen
	\bibfield  {author} {\bibinfo {author} {\bibfnamefont {I.}~\bibnamefont
			{Coddington}}, \bibinfo {author} {\bibfnamefont {W.~C.}\ \bibnamefont
			{Swann}}, \bibinfo {author} {\bibfnamefont {L.}~\bibnamefont {Nenadovic}}, \
		and\ \bibinfo {author} {\bibfnamefont {N.~R.}\ \bibnamefont {Newbury}},\
	}\href {\doibase 10.1038/nphoton.2009.94} {\bibfield  {journal} {\bibinfo
			{journal} {Nature Photonics}\ }\textbf {\bibinfo {volume} {3}},\ \bibinfo
		{pages} {351} (\bibinfo {year} {2009})}\BibitemShut {NoStop}%
	\bibitem [{\citenamefont {Lee}\ \emph {et~al.}(2013)\citenamefont {Lee},
		\citenamefont {Han}, \citenamefont {Lee}, \citenamefont {Bae}, \citenamefont
		{Kim}, \citenamefont {Lee}, \citenamefont {Kim},\ and\ \citenamefont
		{Kim}}]{Lee2013a}%
	\BibitemOpen
	\bibfield  {author} {\bibinfo {author} {\bibfnamefont {J.}~\bibnamefont
			{Lee}}, \bibinfo {author} {\bibfnamefont {S.}~\bibnamefont {Han}}, \bibinfo
		{author} {\bibfnamefont {K.}~\bibnamefont {Lee}}, \bibinfo {author}
		{\bibfnamefont {E.}~\bibnamefont {Bae}}, \bibinfo {author} {\bibfnamefont
			{S.}~\bibnamefont {Kim}}, \bibinfo {author} {\bibfnamefont {S.}~\bibnamefont
			{Lee}}, \bibinfo {author} {\bibfnamefont {S.~W.}\ \bibnamefont {Kim}}, \ and\
		\bibinfo {author} {\bibfnamefont {Y.~J.}\ \bibnamefont {Kim}},\ }\href
	{\doibase 10.1088/0957-0233/24/4/045201} {\bibfield  {journal} {\bibinfo
			{journal} {Measurement Science and Technology}\ }\textbf {\bibinfo {volume}
			{24}} (\bibinfo {year} {2013}),\ 10.1088/0957-0233/24/4/045201}\BibitemShut
	{NoStop}%
	\bibitem [{\citenamefont {Zysk}\ \emph {et~al.}(2007)\citenamefont {Zysk},
		\citenamefont {Nguyen}, \citenamefont {Oldenburg}, \citenamefont {Marks},\
		and\ \citenamefont {Boppart}}]{Zysk2007}%
	\BibitemOpen
	\bibfield  {author} {\bibinfo {author} {\bibfnamefont {A.~M.}\ \bibnamefont
			{Zysk}}, \bibinfo {author} {\bibfnamefont {F.~T.}\ \bibnamefont {Nguyen}},
		\bibinfo {author} {\bibfnamefont {A.~L.}\ \bibnamefont {Oldenburg}}, \bibinfo
		{author} {\bibfnamefont {D.~L.}\ \bibnamefont {Marks}}, \ and\ \bibinfo
		{author} {\bibfnamefont {S.~A.}\ \bibnamefont {Boppart}},\ }\href {\doibase
		10.1117/1.2793736} {\bibfield  {journal} {\bibinfo  {journal} {Journal of
				Biomedical Optics}\ }\textbf {\bibinfo {volume} {12}},\ \bibinfo {pages}
		{051403} (\bibinfo {year} {2007})}\BibitemShut {NoStop}%
	\bibitem [{\citenamefont {Abouraddy}\ \emph {et~al.}(2002)\citenamefont
		{Abouraddy}, \citenamefont {Nasr}, \citenamefont {Saleh}, \citenamefont
		{Sergienko},\ and\ \citenamefont {Teich}}]{Abouraddy2002}%
	\BibitemOpen
	\bibfield  {author} {\bibinfo {author} {\bibfnamefont {A.~F.}\ \bibnamefont
			{Abouraddy}}, \bibinfo {author} {\bibfnamefont {M.~B.}\ \bibnamefont {Nasr}},
		\bibinfo {author} {\bibfnamefont {B.~E.~A.}\ \bibnamefont {Saleh}}, \bibinfo
		{author} {\bibfnamefont {A.~V.}\ \bibnamefont {Sergienko}}, \ and\ \bibinfo
		{author} {\bibfnamefont {M.~C.}\ \bibnamefont {Teich}},\ }\href {\doibase
		10.1103/PhysRevA.65.053817} {\bibfield  {journal} {\bibinfo  {journal}
			{Physical Review A}\ }\textbf {\bibinfo {volume} {65}},\ \bibinfo {pages}
		{053817} (\bibinfo {year} {2002})}\BibitemShut {NoStop}%
	\bibitem [{\citenamefont {Nasr}\ \emph {et~al.}(2003)\citenamefont {Nasr},
		\citenamefont {Saleh}, \citenamefont {Sergienko},\ and\ \citenamefont
		{Teich}}]{Nasr2003}%
	\BibitemOpen
	\bibfield  {author} {\bibinfo {author} {\bibfnamefont {M.}~\bibnamefont
			{Nasr}}, \bibinfo {author} {\bibfnamefont {B.}~\bibnamefont {Saleh}},
		\bibinfo {author} {\bibfnamefont {A.}~\bibnamefont {Sergienko}}, \ and\
		\bibinfo {author} {\bibfnamefont {M.}~\bibnamefont {Teich}},\ }\href
	{\doibase 10.1103/PhysRevLett.91.083601} {\bibfield  {journal} {\bibinfo
			{journal} {Physical Review Letters}\ }\textbf {\bibinfo {volume} {91}},\
		\bibinfo {pages} {8} (\bibinfo {year} {2003})}\BibitemShut {NoStop}%
	\bibitem [{\citenamefont {Lopez-Mago}\ and\ \citenamefont
		{Novotny}(2012)}]{Lopez-Mago2012}%
	\BibitemOpen
	\bibfield  {author} {\bibinfo {author} {\bibfnamefont {D.}~\bibnamefont
			{Lopez-Mago}}\ and\ \bibinfo {author} {\bibfnamefont {L.}~\bibnamefont
			{Novotny}},\ }\href {http://www.ncbi.nlm.nih.gov/pubmed/23027284} {\bibfield
		{journal} {\bibinfo  {journal} {Optics letters}\ }\textbf {\bibinfo {volume}
			{37}},\ \bibinfo {pages} {4077} (\bibinfo {year} {2012})}\BibitemShut
	{NoStop}%
	\bibitem [{\citenamefont {Hong}\ and\ \citenamefont {Mandel}(1985)}]{Hong1985}%
	\BibitemOpen
	\bibfield  {author} {\bibinfo {author} {\bibfnamefont {C.~K.}\ \bibnamefont
			{Hong}}\ and\ \bibinfo {author} {\bibfnamefont {L.}~\bibnamefont {Mandel}},\
	}\href {\doibase 10.1103/PhysRevA.31.2409} {\bibfield  {journal} {\bibinfo
			{journal} {Physical Review A}\ }\textbf {\bibinfo {volume} {31}},\ \bibinfo
		{pages} {2409} (\bibinfo {year} {1985})}\BibitemShut {NoStop}%
	\bibitem [{\citenamefont {Guerreiro}\ \emph {et~al.}(2014)\citenamefont
		{Guerreiro}, \citenamefont {Martin}, \citenamefont {Sanguinetti},
		\citenamefont {Pelc}, \citenamefont {Langrock}, \citenamefont {Fejer},
		\citenamefont {Gisin}, \citenamefont {Zbinden}, \citenamefont {Sangouard},\
		and\ \citenamefont {Thew}}]{Guerreiro2014}%
	\BibitemOpen
	\bibfield  {author} {\bibinfo {author} {\bibfnamefont {T.}~\bibnamefont
			{Guerreiro}}, \bibinfo {author} {\bibfnamefont {a.}~\bibnamefont {Martin}},
		\bibinfo {author} {\bibfnamefont {B.}~\bibnamefont {Sanguinetti}}, \bibinfo
		{author} {\bibfnamefont {J.~S.}\ \bibnamefont {Pelc}}, \bibinfo {author}
		{\bibfnamefont {C.}~\bibnamefont {Langrock}}, \bibinfo {author}
		{\bibfnamefont {M.~M.}\ \bibnamefont {Fejer}}, \bibinfo {author}
		{\bibfnamefont {N.}~\bibnamefont {Gisin}}, \bibinfo {author} {\bibfnamefont
			{H.}~\bibnamefont {Zbinden}}, \bibinfo {author} {\bibfnamefont
			{N.}~\bibnamefont {Sangouard}}, \ and\ \bibinfo {author} {\bibfnamefont
			{R.~T.}\ \bibnamefont {Thew}},\ }\href {\doibase
		10.1103/PhysRevLett.113.173601} {\bibfield  {journal} {\bibinfo  {journal}
			{Physical Review Letters}\ }\textbf {\bibinfo {volume} {113}},\ \bibinfo
		{pages} {173601} (\bibinfo {year} {2014})}\BibitemShut {NoStop}%
	\bibitem [{\citenamefont {Stefanov}(2017)}]{Stefanov2017}%
	\BibitemOpen
	\bibfield  {author} {\bibinfo {author} {\bibfnamefont {A.}~\bibnamefont
			{Stefanov}},\ }\href {\doibase 10.1088/2058-9565/aa6ae1} {\bibfield
		{journal} {\bibinfo  {journal} {Quantum Science and Technology}\ }\textbf
		{\bibinfo {volume} {2}},\ \bibinfo {pages} {025004} (\bibinfo {year}
		{2017})}\BibitemShut {NoStop}%
	\bibitem [{\citenamefont {Teich}\ \emph {et~al.}(2012)\citenamefont {Teich},
		\citenamefont {Saleh}, \citenamefont {Wong},\ and\ \citenamefont
		{Shapiro}}]{Teich2012}%
	\BibitemOpen
	\bibfield  {author} {\bibinfo {author} {\bibfnamefont {M.~C.}\ \bibnamefont
			{Teich}}, \bibinfo {author} {\bibfnamefont {B.~E.}\ \bibnamefont {Saleh}},
		\bibinfo {author} {\bibfnamefont {F.~N.}\ \bibnamefont {Wong}}, \ and\
		\bibinfo {author} {\bibfnamefont {J.~H.}\ \bibnamefont {Shapiro}},\ }\href
	{\doibase 10.1007/s11128-011-0266-6} {\bibfield  {journal} {\bibinfo
			{journal} {Quantum Information Processing}\ }\textbf {\bibinfo {volume}
			{11}},\ \bibinfo {pages} {903} (\bibinfo {year} {2012})}\BibitemShut
	{NoStop}%
	\bibitem [{\citenamefont {Mazurek}\ \emph {et~al.}(2013)\citenamefont
		{Mazurek}, \citenamefont {Schreiter}, \citenamefont {Prevedel}, \citenamefont
		{Kaltenbaek},\ and\ \citenamefont {Resch}}]{Mazurek2013}%
	\BibitemOpen
	\bibfield  {author} {\bibinfo {author} {\bibfnamefont {M.~D.}\ \bibnamefont
			{Mazurek}}, \bibinfo {author} {\bibfnamefont {K.~M.}\ \bibnamefont
			{Schreiter}}, \bibinfo {author} {\bibfnamefont {R.}~\bibnamefont {Prevedel}},
		\bibinfo {author} {\bibfnamefont {R.}~\bibnamefont {Kaltenbaek}}, \ and\
		\bibinfo {author} {\bibfnamefont {K.~J.}\ \bibnamefont {Resch}},\ }\href
	{\doibase 10.1038/srep01582} {\bibfield  {journal} {\bibinfo  {journal}
			{Scientific Reports}\ }\textbf {\bibinfo {volume} {3}},\ \bibinfo {pages}
		{1582} (\bibinfo {year} {2013})}\BibitemShut {NoStop}%
	\bibitem [{\citenamefont {Shirai}\ and\ \citenamefont
		{Friberg}(2014)}]{Shirai2014a}%
	\BibitemOpen
	\bibfield  {author} {\bibinfo {author} {\bibfnamefont {T.}~\bibnamefont
			{Shirai}}\ and\ \bibinfo {author} {\bibfnamefont {A.~T.}\ \bibnamefont
			{Friberg}},\ }\href {\doibase 10.1364/JOSAA.31.000258} {\bibfield  {journal}
		{\bibinfo  {journal} {Journal of the Optical Society of America A}\ }\textbf
		{\bibinfo {volume} {31}},\ \bibinfo {pages} {258} (\bibinfo {year}
		{2014})}\BibitemShut {NoStop}%
\end{thebibliography}
%

\end{document}